\documentclass[a4paper,10pt]{llncs}
\usepackage{color}
\usepackage{balance}
\usepackage{enumerate}
\usepackage{graphicx}
\usepackage[caption=false]{subfig} 
\usepackage{algorithm}
\usepackage{algpseudocode}
\usepackage{amsfonts}
\usepackage{amsopn}
\usepackage{amssymb}
\usepackage{amsmath}
\usepackage{url}
\usepackage{txfonts}
\usepackage{url,epsfig,color}
\usepackage[all]{xy} 
\usepackage[normalem]{ulem}
\usepackage{tabularx}
\usepackage{multirow}
\usepackage{enumitem}

\usepackage{booktabs}

\usepackage[USenglish]{babel}

\newcommand{\matrixX}[1]{\ensuremath{\mathbf{#1}}}

\begin{document}

\title{Collaborative Filtering Ensemble for \\Personalized Name Recommendation}
\titlerunning{Collaborative Filtering Ensemble for Personalized Name Recommendation}

\author{Bernat Coma-Puig\thanks{Work done at the L3S Research Center as part of the ERASMUS exchange\break program while a student at  Universitat Politècnica de Catalunya -- BarcelonaTech (UPC) $<$\texttt{bernat.coma@est.fib.upc.edu}$>$.} \and Ernesto Diaz-Aviles \
and Wolfgang Nejdl}

\authorrunning{Bernat Coma-Puig \and Ernesto Diaz-Aviles \and Wolfgang Nejdl}

\institute{L3S Research Center, Leibniz University Hannover, Germany\\
\{\mbox{coma-puig}, diaz, nejdl\}@L3S.de } \maketitle

\begin{abstract}
Out of thousands of names to choose from, picking the right one for your child is a daunting task. In this work, our objective is to help parents\break making an informed decision while choosing a name for their baby. We follow a\break recommender system approach and combine, in an ensemble, the individual\break rankings produced by simple collaborative filtering algorithms in order to produce a personalized list of names that meets the individual parents' taste.\break Our experiments were conducted using real-world data collected from the query logs of \textit{nameling} (\url{nameling.net}), an online portal for searching and exploring names, which corresponds to the dataset released in the context of the ECML PKDD Discover Challenge 2013. Our approach is intuitive, easy to implement, and features fast training and prediction steps.
\end{abstract}


\section{Introduction}
\label{sec:introduction}
There are many considerations when parents are deciding on a name for their child. Many parents choose to name their babies after a grandparent, other relative, or a close friend. Some others pick names from the actors or actresses of their favorite soap opera. Cultural and societal rules, the meaning of the name, family's traditions, or religious beliefs also play an important role in many countries at the time of choosing a given name for a baby. 

This is indeed a daunting task for the parents and their decision will mark the child for the rest of his or her life. The given name should be unique, making the bearer stand out from the crowd, but at the same time it should also avoid embarrassment of being the source for nicknames, humiliating initials, or annoying email addresses\footnote{such as the one of our friend \textit{H. Thomas Eatons}, who has the (unfortunate) email address of \texttt{eatonsht@<anonymized>.com} :) .}.

From thousands of names to choose from, how do parents pick the right one for their baby? In this paper, we present an approach to help parents dealing with this information overload problem. In particular, we take a recommender systems approach and show how an ensemble of simple collaborative filtering algorithms can help users to find given names that match their needs from a big pool of names.

We conduct this study in the context of the ECML PKDD'13 Discovery Challenge. This paper documents the approach of team ``cadejo'' on the offline phase of the challenge.

The main contribution of this paper is an intuitive approach for the task of given name prediction that is easy to implement, and that features fast training and prediction steps. Our study shows that, in this particular task, our ensemble of simple collaborative filtering building blocks performs significantly better than state-of-the-art latent factor models.


\subsection{Preliminaries}
\label{sec:preliminaries}
We consider the dataset as sparse matrix $\matrixX{X}=[x_{ui}]$, where we use through this paper the letter $u$ for users and $i$ for names, which corresponds to the items in a recommender system setting. We use bold letters for matrices and vectors, and non bold for scalars. The set of users and names\footnote{In this paper, we use the terms ``names" and ``items" interchangeably.} are denoted by $\mathcal{U}$ and $\mathcal{I}$,  respectively. Predictions for user-item pairs are denoted as $\hat{x_{ui}}$. The set of names that the user has interacted with is written as $\mathcal{I}(u)$. The set of users, who interacted with name $i$ is $\mathcal{U}(i)$.

We use the notation $\mathcal{C}_u(i)$ to represent the set of names co-occurring with name $i$ within $\mathcal{I}(u)$, that is, $\mathcal{C}_u(i)  := \{j \mid i,j \in \mathcal{I}(u) \land i \neq j \}$.

We denote the bag of co-occurring names for a given item $i$, as follows: 

\[
	\mathcal{C}(i) :=  \bigcup_{u\in\mathcal{U}} \; \{((i, j), m(i,j)) \mid i \in \mathcal{I}(u) \land j \in \mathcal{C}_u(i)\} \; ,
\]
where $m(i,j): \mathcal{I} \times \mathcal{I} \rightarrow \mathbb{N}_{\geq 1}$ is a function from the set name pairs $(i,j) \in \mathcal{I} \times \mathcal{I}$ to the set $\mathbb{N}_{\geq 1}$ of positive natural numbers. For each pair of names $(i,j)$, $m(i,j)$ represents the number of occurrences of such pair in the bag, i.e., its \textit{multiplicity}. The aggregated bag $\mathcal{C}$ over all items corresponds to $\mathcal{C} := \bigcup _{i \in \mathcal{I}} \mathcal{C}(i)$.

We use $S_u$ to represent the user $u$'s sequence of interactions ordered according to the corresponding timestamp , e.g., if user $u$ searches first for name $\textit{i}_{1}$, after that for $\textit{i}_{4}$ and finally for name $\textit{i}_{2}$, then his sequence $S_u$ is represented as:

\[
	\textit{S}_{u} = \textit{i}_{1} \longrightarrow \textit{i}_{4} \longrightarrow \textit{i}_{2}.
\]

For example, consider three users $\textit{u}_{1}$, $\textit{u}_{2}$ and $\textit{u}_{3}$, and their corresponding sequences $S$ of search actions in temporal order:

\[
\textit{S}_{u1} = \textit{i}_{1} \longrightarrow \textit{i}_{4} \longrightarrow \textit{i}_{2} \longrightarrow  \textit{i}_{3}
\]
\[
\textit{S}_{u2} = \textit{i}_{4} \longrightarrow \textit{i}_{5} \longrightarrow \textit{i}_{1} \longrightarrow  \textit{i}_{4} \longrightarrow  \textit{i}_{3}
\]
\[
\textit{S}_{u3} = \textit{i}_{3} \longrightarrow \textit{i}_{5} \longrightarrow \textit{i}_{6} \longrightarrow  \textit{i}_{7} \longrightarrow  \textit{i}_{4}
\]
\newpage
The bag of co-occurrences for item $\textit{i}_{4}$, $\mathcal{C}(i_4)$, sorted in decreasing order of multiplicity, is given by:

\[
 \mathcal{C}(i_4) = \{((i_4,i_3),3), ((i_4,i_1),2), ((i_4,i_5),2), ((i_4,i_2),1), ((i_4,i_6),1), ((i_4,i_7),1)\} \; .
\]

\subsection{The Dataset}
\begin{figure}[!hb]
        \centering
        \subfloat[][\textbf{Frequency of names per given user.}]{
                \centering
		\includegraphics[width=0.5\textwidth]{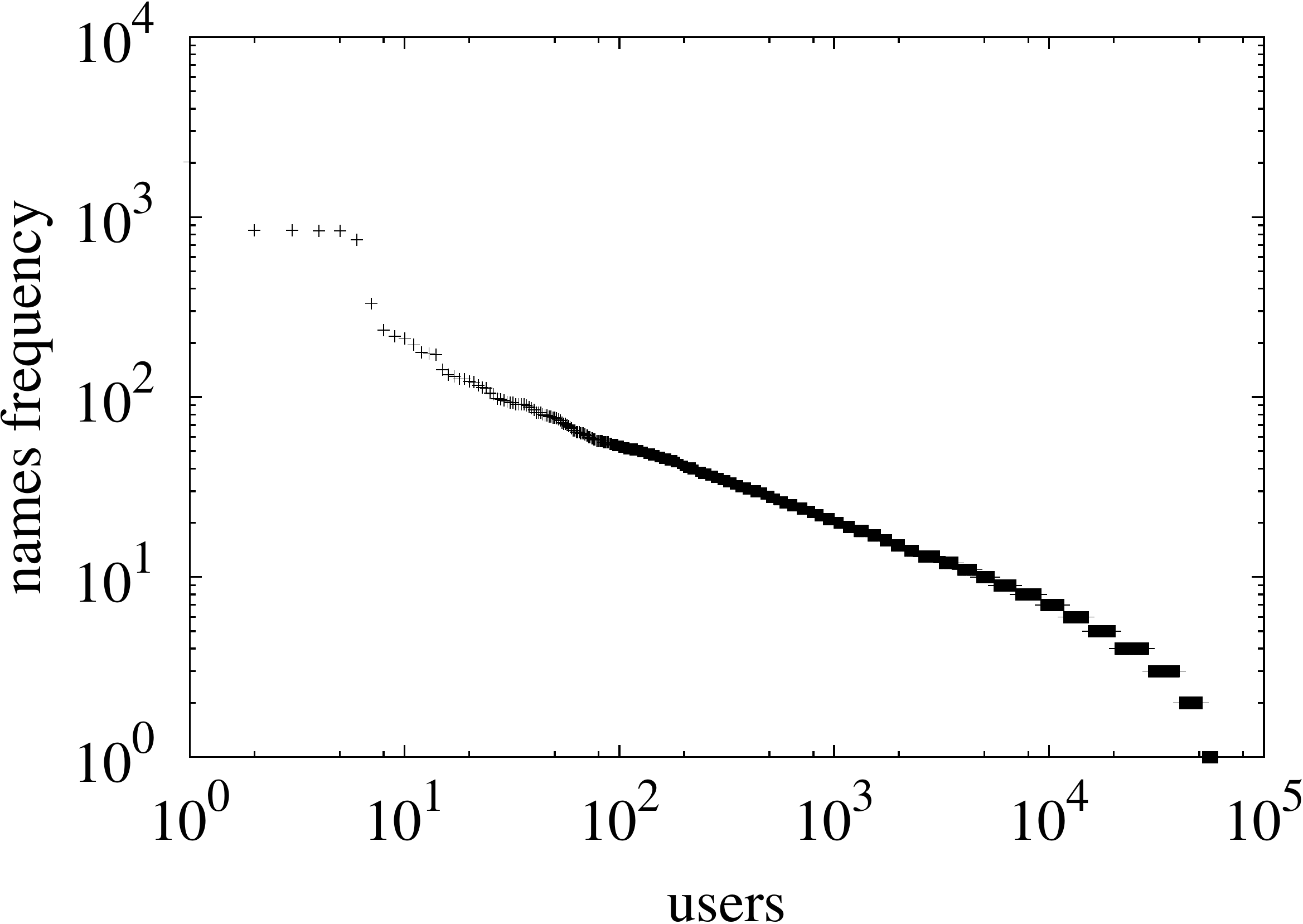}
		\label{fig:users_freq}
	}
        ~ 
        \subfloat[][\textbf{Frequency of users per given name.}]{
                \centering
		\includegraphics[width=0.5\textwidth]{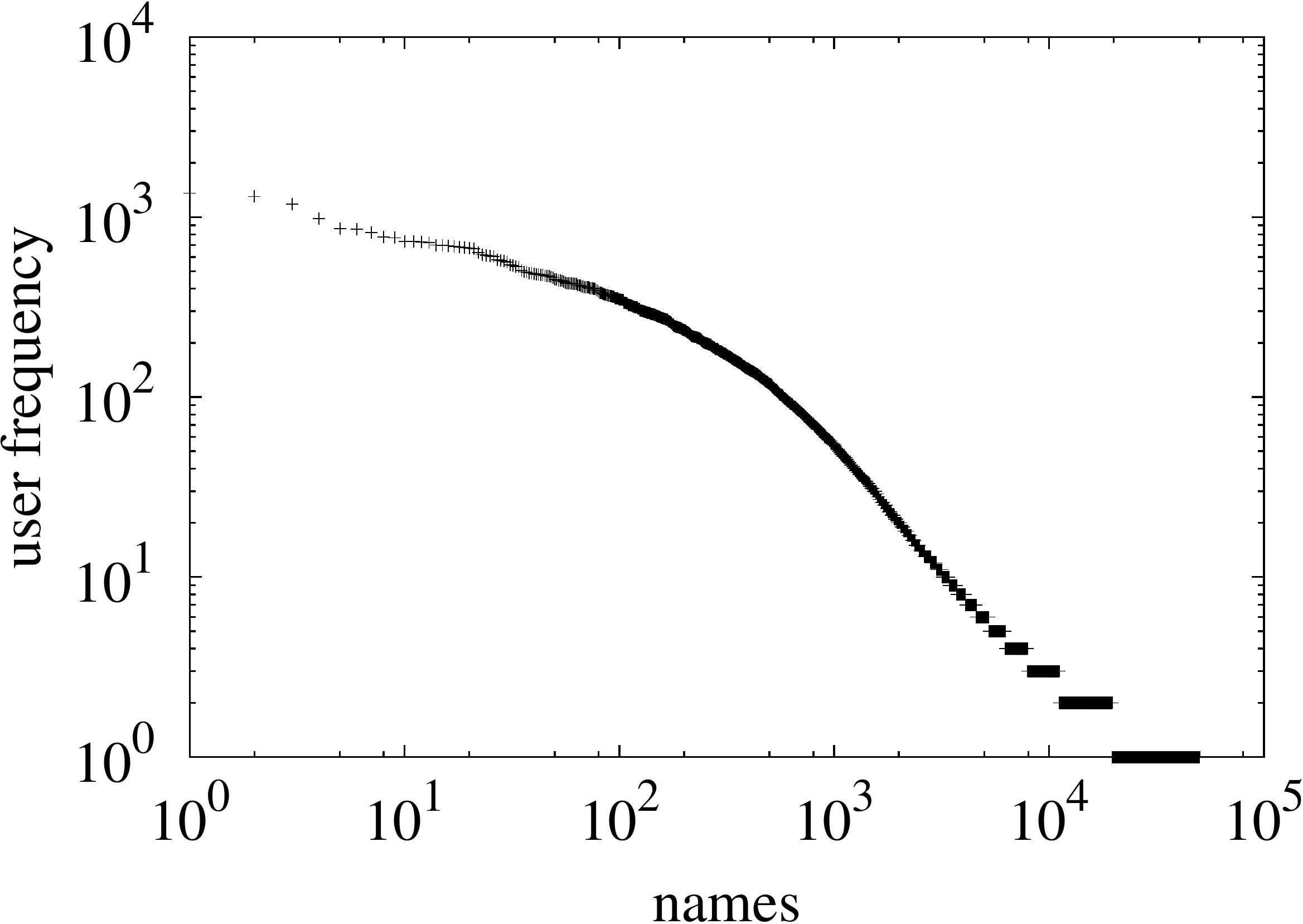}
		\label{fig:names_freq}
	}
        \caption{\textbf{Frequency of users and names.}}
	\label{fig:power_law}
\end{figure}

The dataset provided for the offline challenge is based on the query logs of \textit{nameling} (\texttt{nameling.net}), an online portal for searching and exploring names. The collection comprises the time period from March 6th, 2012 to February 12th, 2013. In total the dataset contains 515,848 interactions (i.e., activities) from $60,922$ different users and $50,277$ unique names. Figure~\ref{fig:users_freq} shows the frequency of names per user. We can observe that it corresponds to a characteristic graph of a long-tail distribution, where few names tend to concentrate a large number of users. The frequency of users per given name is shown in Figure~\ref{fig:names_freq}.

There are 5 different types of interactions within the dataset, which are described as follows:

\begin{enumerate}
	\item ENTER\_SEARCH: the user explicitly writes a name in the search field of nameling's website in order to search for it.

	\item LINK\_SEARCH: the user clicks on a name of showed names at nameling's website, following a link to a search result page.

	\item NAME\_DETAILS: the user requests more detailed information of a name.

	\item LINK\_CATEGORY\_SEARCH: Wherever available, a name is categorized according to the corresponding Wikipedia article. Users may click on such a category link to obtain all names in the corresponding category.

	\item ADD\_FAVORITE: the user saves the name in his list of favorite names.
\end{enumerate}

In addition to these datasets there is a list of valid or \textit{known names} provided by the organizers of the challenge, which contains 36,436 given names.

\subsection{The Task}
The task for the offline challenge is to recommend a personalized ranked list of names for each user in the test set, based on the users' (partial) search history in nameling. 

The recommender system's quality is evaluated with respect to the set of names that users have entered directly into nameling's search field. The rationale to restrict the evaluation to ENTER\_SEARCH activities is that all other user activities are biased towards the lists of names which were displayed to nameling users. 

The test set is built by taking from the training data the chronologically last two names, which had directly been entered into nameling's search field (i.e., using the ENTER\_SEARCH activity) and which are also contained in the list of known names as detailed in the challenge description\footnote{\url{http://www.kde.cs.uni-kassel.de/ws/dc13/faq/} .}.

The assessment metric for the recommendations is Mean Average Precision at a \mbox{cut-off} of 1000 (MAP@1000)~\cite{baeza2011}. That is, for each test user look up the left-out names and take the precision at the respective position in the ordered list of recommended names. These precision values are first averaged per test user and than in total to yield the final score. MAP@1000 means that only the first 1,000 positions of a list are considered. Thus it might happen that for some test users one or both of the left out names do not occur in the list of recommendations. These cases will be handled as if the missing names were ranked at position 1001 and 1002 respectively.

\subsection{Data Preprocessing and Validation Set}
In our study we could not find a clear mechanism on how to exploit activities of type LINK\_CATEGORY\_SEARCH, and therefore we drop such interactions from the dataset. We also concentrate only on names which appear as part of at least one interaction and that were also present in the known names list. In total our experiments consider a total number of 260,236 user-name pair interactions, from $|\mathcal{U}|=60,922$ different users and $|\mathcal{I}|=17,467$ unique names. The mean of names per user is $4.35$, the median is $3$ names per user, with a minimum $1$ and a maximum of $1670$ names per user.

To evaluate our results we built a cross-validation dataset using the script provided by the organizers of the challenge. The script gives us a validation with 13,008 users and two target names. From these 13,008 users, 2,264 are not within the 4,140 users in the test set, which are the ones we are required to give recommendations. In order to have a more representative cross-validation dataset, for each of these 2,264 users we also selected, from the remaining transactions in the training set, the last 2 names the user interacted with. Note that in this case we ignored the additional constraints imposed by the script, e.g., the type of activity.

\section{Related Work}
\label{sec:relatedwork}
Although top-$N$ recommender systems have been studied in depth, the particular task of recommending given names is rather new. For example recent work by Mitzlaff et al. studies the relatedness of given names based on data from the social web~\cite{mitzlaff2013onomastics}.\break This work shows the importance of co-occurrence networks for the recommendation task. Our approach also exploits name co-occurrences in the Name-to-Name algorithm introduced in Section~\ref{sec:methods}.

The NameRank algorithm introduced in~\cite{mitzlaff2013recommending} adapts FolkRank~\cite{DBLP:conf/esws/HothoJSS06} for name recommendation, showing promising results. The algorithm basically solves a personalized version of PageRank~\cite{Brin:1998:ALH:297805.297827} per user in the system, over a graph of names, which does not scale gracefully to large-scale data. Our approach, on the other hand, is flexible enough to combine multiple predictors from simple collaborative filtering models, which makes it more attractive for big data scenarios.

\section{Methods}
\label{sec:methods}
Collaborative Filtering (CF) algorithms are best known for their use on e-commerce Web sites, Online Social Networks, or Web 2.0 video sharing platforms, where they use input about a user's interests to generate a (ranked) list of recommended items. In this section, we describe the collaborative filtering models which are used by our approach as well as the assembling strategy to compute the final prediction.

\subsection{Name-to-Name Collaborative Filtering}
This approach for name recommendation is based on the classic item-based collaborative filtering algorithm introduced by Amazon.com~\cite{1167344}. This algorithm matches each user's interaction with a name to a set of similar names, then combines those similar items into a recommendation list. 

To determine the most similar match for a given name, the algorithm constructs a bag of co-occurring names across all user interactions in the collection. The rationale behind this algorithm is that there are many names that do not co-occur in any of the user's name transactions ($\mathcal{I}(u)$), and thus the approach is efficient in terms of processing time in memory, since there is no need to compute similarities over all possible pairs of names in the collection.

To compute the final recommendation list, the algorithm finds names similar to each of the ones in the user's set of names $\mathcal{I}(u)$, aggregates those co-occurring names, and then recommends the most popular or correlated names. This computation is very quick, depending only on the number of names the user has interacted with. 

The Name-to-Name algorithm is summarized in Algorithm~\ref{alg:name2nameCF}\footnote{In Section~\ref{sec:results}, we explain the models used in our ensemble and provide more details about Algorithm~\ref{alg:name2nameCF}'s  functions $\texttt{getRandomName}()$ and $\texttt{getRandomCoName}()$ in this context.}.

\begin{algorithm}[!htb]
\renewcommand{\algorithmicrequire}{\textbf{Input:}}
\renewcommand{\algorithmicensure}{\textbf{Output:}}
\caption{\textsc{Name-to-Name CF}}
\label{alg:name2nameCF}
\begin{algorithmic}[1]
\Require {
\Statex Target user $u \in \mathcal{U}$. Recommendations will be computed for this user;
\Statex $\mathcal{I}(u) \subset \mathcal{I}$: set of names that the user has interacted with;
\Statex $\mathcal{C}$ : bag of co-occurring names;
\Statex $N \in \mathbb{N}$: size of the recommendation list; 
\Statex max\_iterations: maximum number of iterations.
\Ensure {Recs($u$): ranked list of recommendations for user $u$.}
\Statex
\Procedure{\textsc{GetRecommendations}}{$u$, $\mathcal{I}(u)$, $\mathcal{C}$, $N$, max\_iterations}
    \State Recs($u$) $\gets \emptyset$
    \State Recs($u$) $\gets$ \textsc{Name-to-Name}($u$, $\mathcal{I}(u)$, $\mathcal{C}$, $N$, Recs($u$), max\_iterations)
    \While{$|$Recs($u$)$| < N$}
	\State Recs($u$) $\gets$ \textsc{Name-to-Name}($u$, Recs($u$), $\mathcal{C}$, $N$, Recs($u$), max\_iterations)
    \EndWhile
    \State\Return{\texttt{sort}(Recs($u$))}
\EndProcedure
\Statex
\Procedure{\textsc{Name-to-Name}}{$u$, $\mathcal{I'}(u)$, $\mathcal{C}$, $N$}
    \While{$|$Recs($u$)$|$ $< N$ \textbf{and} $t < $ max\_iterations}
        \State $i \gets $\texttt{getRandomName}($\mathcal{I'}(u)$)
	\State $((i,j),m) \gets $\texttt{getRandomCoName}($\mathcal{C}(i)$)
	\If{$j \not\in $ Recs($u$) \textbf{and} $j \not\in$ $\mathcal{I'}(u)$}
		\State Recs($u$) $\gets$ Recs($u$) $\cup$ $ \{(j,m)\}$ 
	\EndIf
	\State $t \gets t + 1$
    \EndWhile
    \State\Return{Recs($u$)}
\EndProcedure
}
\end{algorithmic}
\end{algorithm}

\subsection{Neighborhood-based Collaborative Filtering}
Neighborhood-based recommendation is a classic approach for Collaborative Filtering that still ranks among the most popular methods for this
problem. These approaches are quite simple to describe and implement featuring important advantages such as the ability to explain a recommendation and to capture ``local'' relations in the data, which are useful in making serendipitous recommendations. In particular, we used the Top-N variants of the User-Based and Item-Based algorithms~\cite{DBLP:reference/rsh/DesrosiersK11} as part of our name recommendation ensemble.

\subsection{Ensemble}
\label{sec:ensemble}
Our solution to the challenge consists of an ensemble of individual rank estimates of a set of collaborative filtering algorithms, a method that has shown to improve the quality of the recommendations~\cite{Bell_2007}.

Since the value estimates of our models, $\hat{x}$, can be in different scales, we do not combine their values directly, but rather we use their rank estimates. Formally, the ensemble of the rank estimates of $l$ models is given by:

\begin{equation}
	\hat{x}^{\text{rank}}_{ui} := \sum_{l}{\alpha_l \cdot \frac{1}{\text{rank}(\hat{x}^l_{ui})}} \; ,
\label{eq:ensemble}
\end{equation}
where $\alpha_l$ is a weight associated to the predictors of model $l$, $\text{rank}(\hat{x}^l_{ui})$ is the rank position within the $l$th ranked list corresponding to the estimate value $\hat{x}^l_{ui}$. That is, the combined estimate $\hat{x}^{\text{rank}}_{ui}$ corresponds to the weighted reciprocal rank of the individual models.

\section{Results}
\label{sec:results}
In this section, we detail the collaborative filtering models, report their parameters, and individual recommendation performance in terms of MAP@1000. We also present the performance boost achieved by our ensemble.

The ensemble of our solution consists of 9 collaborative models that we describe as follows.

\begin{description}
	\item[\textbf{[m0 -- N2N-Freq]}] is a Name-to-Name CF model that is created using the names co-occurring with the names of a given test user, according to Algorithm~\ref{alg:name2nameCF}.
This model considers the ``ENTER\_SEARCH'', ``LINK\_SEARCH'' and ``NAME\_DETAILS'' activites to build the bag of co-occurrences. We randomly select a name $i$ for given test user $u$ via the \texttt{getRandomName($\cdot$)} procedure specified in Algorithm~\ref{alg:name2nameCF}, where the chance for a name to be chosen is proportional to how often user $u$ has interacted with it, which adds a positive bias towards those names that are more searched by the user. Furthermore, we also bias the selection of the co-occurring name $j$ (\texttt{getRandomCoName($\cdot$)} procedure in Algorithm~\ref{alg:name2nameCF}) towards the multiplicity of the pair $(i, j)$.

\noindent\textit{Example.} To illustrate this approach, consider the following example. Our dataset consists of five users, $u_1 \ldots u_5$, and our task is to predict a recommendation list of names for user $u_1$. The sequence of interactions for user $\textit{u}_{1}$ is denoted as $S_{u1}$ (cf. Section~\ref{sec:preliminaries}) is given by
\[
	\textit{S}_{u1} = \textit{i}_{4} \longrightarrow \textit{i}_{1} \longrightarrow \textit{i}_{4}.
\]
and for the other four users, their corresponding sequences are:

\[
\textit{S}_{u2} = \textit{i}_{1} \longrightarrow \textit{i}_{4}  \longrightarrow  \textit{i}_{3}
\]
\[
\textit{S}_{u3} = \textit{i}_{4} \longrightarrow \textit{i}_{5} \longrightarrow \textit{i}_{1} \longrightarrow  \textit{i}_{4} \longrightarrow  \textit{i}_{3}
\]
\[
\textit{S}_{u4} = \textit{i}_{3} \longrightarrow \textit{i}_{6} \longrightarrow  \textit{i}_{7} \longrightarrow  \textit{i}_{4}
\]
\[
\textit{S}_{u5} = \textit{i}_{1} \longrightarrow \textit{i}_{5} \longrightarrow \textit{i}_{2}
\]
then, the bags of co-occurrences for the names in $\textit{S}_{u1}$, i.e., $\textit{i}_{4}$ and $\textit{i}_{1}$, sorted in decreasing order of multiplicity, are given by:

\[
 \mathcal{C}(i_4) = \{ ((i_4,i_1),3), ((i_4,i_3),3), ((i_4,i_5),1), ((i_4,i_6),1), ((i_4,i_7),1)\}  .
\]
\[
 \mathcal{C}(i_1) = \{((i_1,i_4),3), ((i_1,i_3),2), ((i_1,i_5),2),  ((i_1,i_2),1)\}  .
\]

Using the N2N-Freq shown in Algorithm~\ref{alg:name2nameCF}, we first chose one name from user $u_1$'s names (i.e., from $\mathcal{I}(u_1)$), and the name's corresponding bag of co-occurrences. Let us assume that $i_4 \in \mathcal{I}(u_1)$ and its respective bag $\mathcal{C}(i_4)$ are chosen.\break The first item to be included in the list of recommendations is $\textit{i}_{3}$ ($\textit{i}_{1}$ would not be chosen because $i_1 \in \mathcal{I}(u_1)$). 

In the next iteration, consider that $\mathcal{C}(i_4)$ is selected again, given that it has a higher probability to be picked due to the frequency of item $i_4$ in the sequence $S_{u1}$. In this case, $\textit{i}_{5}$ would be the item chosen to be included in the list of recommendations. 

In the third iteration, the list selected is $\mathcal{C}(i_1)$, then the first item to be selected is $\textit{i}_{2}$. Note that there are no more items from $\mathcal{C}(i_1)$ that can be included in the list. Then, $R(\textit{u}_{1})$ is filled up using items from $\mathcal{C}(i_4)$. 

Finally, the list of recommendations for $\textit{u}_{1}$ corresponds to:
\[
R(\textit{u}_{1}) = [\textit{i}_{3}, \textit{i}_{5}, \textit{i}_{2}, \textit{i}_{6}, \textit{i}_{7}] \; .
\]
	\item[\textbf{[m1 -- N2N-Freq-ES]}] follows the same approach as model m0, but the bag of co-occurring names used to compute the predictions considers only the\break``ENTER\_SEARCH'' activity to build the bag of co-occurrences.
\\\hfill
	\item[\textbf{[m2 -- N2N-Time]}] is also a Name-to-Name CF model similar to m0, but with the difference that the names $\mathcal{I}(u)$ are not selected biased towards frequency of user interactions, but towards recency. That is,  the names included in the recommendation list are those that co-occur with the last searches of the test user. The goal of this model is to capture the latest user preferences as input to compute the recommendations.
\\\hfill
\noindent\textit{Example.} Using this algorithm, with $\textit{S}_{u1}$, $\textit{S}_{u2}$, $\textit{S}_{u3}$, $\textit{S}_{u4}$ $\textit{S}_{u5}$, $\mathcal{C}(i_4)$ and $\mathcal{C}(i_1)$ from the example given for \textbf{m0}. Using this algorithm, biased towards recency, all selectable items from $\mathcal{C}(i_4)$ have a higher probability of being chosen. The firsts items would correspond to $\textit{i}_{3}$, $\textit{i}_{5}$ and $\textit{i}_{6}$. From $\mathcal{C}(i_1)$ the selectable items are $\textit{i}_{7}$ and $\textit{i}_{2}$. A possible recommendation list corresponds to:  
\[
R(u_{1}) = [i_{3}, i_{5}, i_{6}, i_{7}, i_{2}] \; .
\]
	\item[\textbf{[m3 -- N2N-Time-ES]}] follows the same temporal strategy as m2, but the bag of co-occurring names only considers the ``ENTER\_SEARCH'' activity.
\\\hfill
	\item[\textbf{[m4 -- N2N-Time-NoTop5]}] this model is the same one as m2, but only the top-5 most popular names are excluded from the bag of co-occurrences. The rationale behind this model is to get a more specific list of names, avoiding the names that are too popular.
\\\hfill
	\item[\textbf{[m5 -- N2N-Time-NoTop10]}] This model is similar to model m2, with the exception that the top-10 most popular names in the collection have not been considered to build the bag of co-occurrences.
\\\hfill
	\item[\textbf{[m6 -- UB-T]}] is a user-based collaborative filtering algorithm~\cite{DBLP:reference/rsh/DesrosiersK11} using Tanimoto coefficient for binary feedback as similarity metric~\cite{tanimoto}. We used a neighborhood of size 100\footnote{Observe that we did not optimize for this parameter.}.
\\\hfill
	\item[\textbf{[m7 -- UB-LL]}] is a user-based model that uses likelihood as similarity metric. As in the previous model, we also used a neighborhood of size 100 in this case.
\\\hfill
	\item[\textbf{[m8 -- PR]}] This model corresponds to PageRank~\cite{Brin:1998:ALH:297805.297827} computed on the graph of co-occurring names. This is a non-personalized recommendation algorithm biased to the most popular items. We used this algorithm to ``fill up'' recommendation lists with less than 1000 names per user. 

\end{description}

All models, except m6 and m7, were implemented in the Python programming language, using the numeric libraries of NumPy and SciPy\footnote{\url{http://www.scipy.org/} .}. For the user-based models (m6 and m7), we used the Java implementation provided by Apache Mahout\footnote{\url{http://mahout.apache.org/} .}. 
\\\hfill

Table~\ref{table:map} summarizes the individual performance of these models. We also report the performance of a non-personalized model that always recommends the most popular 1000 names.

\begin{table}[!t]
\centering
\scalebox{1}{
\begin{tabular}{c l r}
\toprule
\textbf{Model} & \textbf{Description} & \textbf{MAP@1000}\\
\midrule
m0  & N2N-Freq & 0.033449\\
m1  & N2N-Freq-ES & 0.033430\\ 
m2  & N2N-Time & 0.032296\\
m3  & N2N-Time-ES & 0.032008\\
m4 & N2N-Time-NoTop5 & 0.032526\\
m5  & N2N-Time-NoTop10 & 0.032455\\
m6  & UB-T & 0.023921\\
m7  & UB-LL & 0.028365\\
m8 & PR & 0.026483\\
\midrule
& \textbf{Final ensemble} &  \textbf{0.036766} \\
\midrule
baseline & Most Popular Names & 0.028138\\
\bottomrule
\\
\end{tabular}
}
\caption{\textbf{Recommendation performance in terms of MAP@1000 for the individual models and the final ensemble. The performance of a non-personalized model that always recommends the most popular 1000 names is reported as baseline.}}
\label{table:map}
\end{table}

\subsection*{Engineering the Final Ensemble}
\begin{figure}[!ht]
\centering
\includegraphics[width=\textwidth]{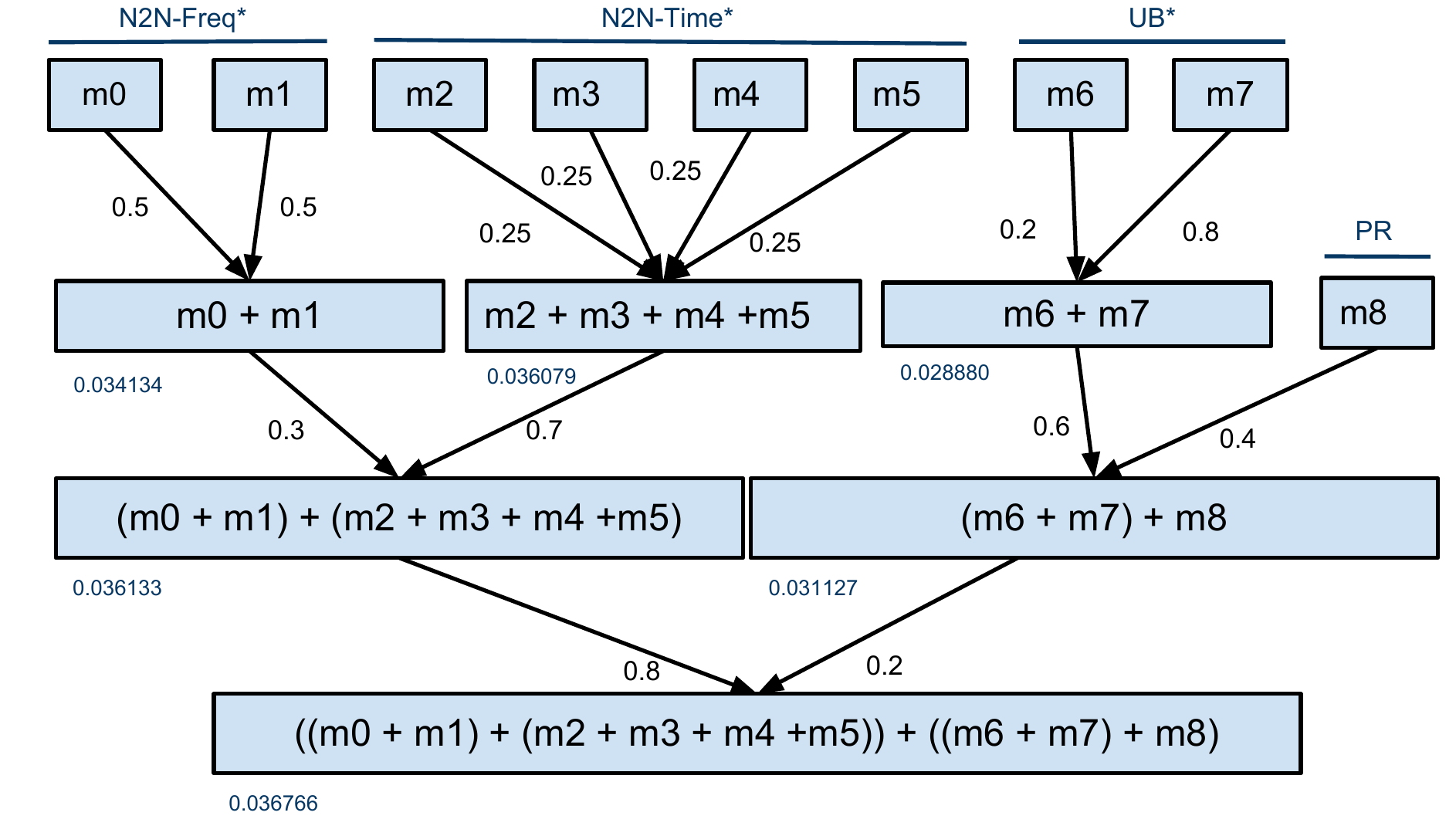} 
\caption{\textbf{Final ensemble. The $\alpha$ weights for the partial model ensembles are indicated next to the corresponding arrow. The symbol `+' indicates the assembling of the models. The MAP@1000 for the corresponding sub-ensembles are shown below the respective boxes.}}
\label{fig:ensemble}
\end{figure}

We compute the final ensemble by first combining different \textit{flavors} of the same approach, and then combining the resulting ranked lists as explained in Section~\ref{sec:ensemble}. Figure~\ref{fig:ensemble} illustrates the assembling process.

All weights (the $\alpha$'s in Equation~\ref{fig:ensemble}) were determined experimentally based on the performance achieved by the (sub-)ensambles in our cross validation set.
We found that the best way to combine the N2N-Freq* (m0 and m1) and N2N-Time* (m2 and m3) algorithms was by giving them equal weights, this is not surprising given their very similar performance. On the other hand, the performance of the User Based algorithms differs more substantially. In this case, we found that the best way to combine them was by giving a higher weight to UB-LL ($\alpha_{UB-LL} = 0.8$) and a weight of $\alpha_{UB-T}=0.2$ to UB-T, for a UB combination (m6 + m7) that achieved a MAP@1000 of 0.028880.\break We combine the unpersonalized ranked list output by the PageRank (m8) with the UB ensemble to fill up user's lists with less than 1000 items, using an asymmetric weighting scheme, favoring the UB combination.

The final ensemble combines the N2N family combinations with the ranked list from the filled UB models. We found that the best combination was obtained by giving the N2N and UB*+PR a weights of 0.8 and 0.2, respectively. The MAP@1000 for the final ensemble reaches a value of 0.036766. Please refer to Table~\ref{table:map} to compare the ensemble's performance to the one of the individual models.

\section{General Thoughts}
\label{sec:discussion}
The low values of MAP@1000 obtained by our approach on this dataset give an idea of how difficult the problem of recommending given names is. 

Given the evaluation design of hiding the last two names the user interacted with, models that capture the latest user preferences, e.g., from the session information, tend to work well for us. 

Neighborhood-based algorithms perform worse than item-to-item co-occurrences. Within the item-based and user-based variants, we observed that results from item-based collaborative filtering were inferior to the ones achieved by the user-based models, and therefore we did not consider them in the ensemble.

One of CF's most successful techniques are low dimensional linear factor models, that assume user preferences can be modeled by only a small number of latent factors. One of such methods is matrix factorization, which has been effectively used for the rating and item prediction task~\cite{Koren:2009:MFT:1608565.1608614}. 

We conducted extensive experiments using state-of-the-art CF algorithms based on matrix factorization. In particular, we evaluated the performance of BPR~\cite{Rendle:2009:BBP:1795114.1795167} and RMFX~\cite{Diaz-Aviles:2012:RTR:2365952.2365968} for the challenge's task, but we found that the performance achieved was only at the level of a baseline predictor that recommends the most popular names. This poor performance of matrix factorization models has been also observed by Folke et al.~\cite{mitzlaff2013recommending}.

We also learned a name-to-name similarity matrix from the co-occurring names adjacency via optimizing a ranking criteria, as suggested in~\cite{Rendle:2009:BBP:1795114.1795167}, the results were also discouraging.

Furthermore, we also tried to optimize directly for MAP following a Learning to Rank framework suggested in~\cite{Diaz-Aviles:2012:SRR:2365952.2366001} and \cite{Balakrishnan:2012:CR:2124295.2124314}. This approach learns the latent factors for users and items, and then applies standard Learning to Rank to optimize for a desired metric. Our results did not reach the level of the baseline predictor of most popular names.

Given this performance, we did not include any latent factor model in our ensemble. Why the results achieved using latent factor models, for this particular task of name prediction, are inferior to the ones obtained with simple methods? In our future research, we plan to explore this question more in detail.

\section{Conclusion}
\label{sec:conclusion}
In this paper, we presented an ensemble of several algorithm for personalized ranked recommendation of given names. We found that the co-occurring name information was a key component for the Name-to-Name algorithms used in our ensemble. Our method is intuitive and simple to implement, and does not suffer from the scalability issues as previous methods introduced for this task.

As a future work, we plan to further explore this interesting challenge in order to help parents deciding what is the best name for their baby.
%
%

\section*{Acknowledgements}
We would like to thank Asmelash Teka and Rakshit Gautam for their valuable feedback. This work is funded, in part, by the L3S IAI research grant for the \textit{FizzStream!} Project. Bernat Coma-Puig is sponsored by the EuRopean Community Action Scheme for the Mobility of University Students (ERASMUS).

\bibliographystyle{abbrv}
\bibliography{biblio}  
\end{document}